\documentclass[aps,prc,twocolumn]{revtex4}
\usepackage{amsmath}

\usepackage{graphicx}
\usepackage{dcolumn}
\usepackage{bm}

\begin{document}


\title{Hamilton's principle: why is the integrated {\em difference}  of
kinetic and potential energy minimized?}


\author{Alberto G. Rojo}
\email{rojo@oakland.edu}
\affiliation{%
Department of Physics, Oakland University, Rochester, MI 48309.
}%


\date{\today}

\begin{abstract}
I present an intuitive answer to an often asked question: why is
the integrated {\em difference} $K-U$ between the kinetic and
potential energy  the quantity to be minimized in Hamilton's
principle?
 Using elementary arguments, I map
the problem of finding the path of a moving particle connecting
two points to that of finding the minimum potential energy of a
static string. The mapping implies that the configuration of a
non--stretchable string of variable tension corresponds to the
spatial path dictated by the Principle of Least Action; that of a
stretchable string in space-time is the one dictated by Hamilton's
principle. This correspondence  provides the answer to the
question above: while a downward force curves the trajectory of a
particle in the $(x,t)$ plane downward,  an upward force of the
same magnitude stretches the string to the same  configuration
$x(t)$.

\end{abstract}

\maketitle

\section{\label{sec:level1} Introduction}

 Nature loves extremes. Soap films seek to minimize
their surface area and adopt a spherical shape; a large piece of
matter tends to maximize the gravitational attraction between its
parts and, as a result, planets are also spherical; light rays
refracting on a magnifying glass bend and follow the path of least
time, and a relativistic particle chooses to follow the path
between two events in space-time that maximizes the time measured
by a clock on the particle.\cite{taylor1,foot0}   Nor are humans
divorced from extremum principles, attempting to reduce the
complexities of the world to the smallest number of dominating
principles.

 For mechanical systems, this goal has been
accomplished by Maupertuis and Hamilton in different  versions of
the celebrated Principle of Least Action. Both formulations emerge
from conceiving the trajectories of particles  as light rays
traversing different media which, according to Fermat, follow the
path that minimizes time.

In 1744,  Maupertuis proposed that ``Nature,
 in the production of its effects, does so always by the simplest means"
 \cite{maupertuis1} and, in 1746, wrote ``in Nature, the action ({\em la quantit{\'{e}} d'action})
 necessary for change is the smallest possible. Action is the product of
 the mass of a body times its velocity times the distance it moves".\cite{maupertuis2}
 For a light ray or a particle  passing from
one medium into another, both the minimization of time and
minimization of action gives rise to angles of incidence and
refraction in a fixed proportion to each other: the analog of
Snel's law\cite{snel}  for a light ray $n_1\sin \theta_1=n_2\sin
\theta_2$ corresponds to the conservation of particle momentum
along the interface $mv_1\sin \theta_1=mv_2\sin \theta_2$. Even
though Maupertuis' formulation was vague--and there was
controversy on the priority over the idea\cite{maupertuis3}--his
name remains attached to the principle arguably for two reasons:
his metaphysical view that minimum action expresses God's wisdom
in the form of an economy principle\cite{terrall}, and Euler's
role in settling the controversy in his favor.

Hamilton's principle, formulated almost a century later, is
similar to the Principle of Least Action and is based on the
optical-mechanical analogy as well.\cite{hamilton} While the
trajectory followed by a particle of fixed energy $E$ connecting
two points in space is prescribed by the Principle of Least
Action, Hamilton's principle determines the trajectory for which
the particle will spend a given time $t$ in travelling between the
same points. In such a case, the optimal path is the one for which
the sum of the products $(K-U)\Delta t$ along the path is a
minimum ($K$ and $U$ are the kinetic and potential energies and
$\Delta t$ the time interval). Hamilton's method was mentioned
throughout the nineteenth century but was rarely used in practice
because simpler methods were as effective in most
cases.\cite{hamilton1}

The situation changed in 1926 when Schr\"{o}dinger, inspired by de
Broglie's ideas, resorted to Hamilton's analogy between mechanics
and geometric optics, extended the treatment from geometrical to
undulatory mechanics, and arrived at his famous equation for the
dynamics of a quantum mechanical particle.\cite{schrodinger}
Moreover, in 1948, Feynman\cite{feynman} (following a hint from
Dirac) offered a new perspective on Hamilton's principle: in
propagating between two points, a quantum particle ``explores"
every path, treating them on equal footing. This ``democracy of
histories"\cite{mtw} becomes the Principle of Least Action for a
classical particle due to destructive interference that eliminates
those paths that differ significantly from the classical (or
extremal) path.

\begin{figure}
\includegraphics*[width=0.5\textwidth]{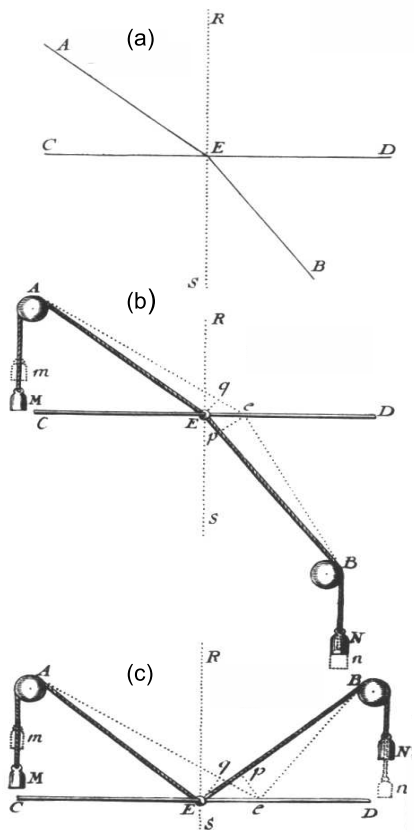}
\vspace{-0.5cm}

 \caption{(a) and (b) John Bernoulli's proof of Snel's law
using the mechanical equilibrium of a tense string. (c) Mechanical
model for the law of reflection, which is not discussed in the
present paper.  (Figure reproduced from Reference
\cite{bernoulli}). }
 \label{figber}
\end{figure}

A lesser known approach to what is called now the Principle of
Least Action was taken by John Bernoulli\cite{bernoulli}, who
showed that Snel's law can be obtained from the condition of
mechanical equilibrium of a tense, non-stretchable string (see
Figure \ref{figber}). This analogy was also noted by
M\"{o}bius\cite{mobius0,mobius} and discussed by Ernst Mach in his
classical book on mechanics.\cite{mach} Ref. \cite{cebreros}
considers an inextensible string and is the only article I was
able to find on this analogy.

 In the present paper I show that a
simple extension of this idea to paths that are covered in fixed
time can be used to prove the equivalence of Hamilton's principle
to the static equilibrium of a {\em stretchable} string. My aim is
to provide, in the spirit of References
\cite{hanc,moore,hanc2,hanc3,hanc4}, insight into why it is the
 the difference between the kinetic and
potential energy that appears in Hamilton's principle. In section
\ref{unststr}, I review Bernoulli's approach and in section
\ref{hamnc}, I present a simplified derivation of Hamilton's
principle without calculus. In section \ref{hamc1}, I present a
slightly more elaborate derivation using elementary calculus.
Given the importance of the Principle of Least Action  in so many
areas of Physics, I hope that this paper will contribute to its
presentation in introductory courses rather than it being
postponed to advanced mechanics courses.

\section{\label{unststr} The Least Action Principle and non-stretchable strings}

Figure \ref{figber} shows the diagram used by John Bernoulli to
derive Snel's law for a light ray travelling from point $A$ to $B$
 using the analogy with the static equilibrium of a tense string.
 The following is a  rephrasing of Bernoulli's argument, which
 is based on the hypothesis that, in mechanical equilibrium,
 it is equivalent
 to say that the net force on each point of the system is zero,
 and that the system is in the state of minimum potential energy.
 In discussing the particle motion I will assume knowledge of Newton's law
 $\textbf{F}=\Delta \textbf{p}/\Delta t$
 relating the force on a particle with the rate of chance of its
 momentum.

  Call $T_1$ and $T_2$ the weights hanging from points  $A$ and
$B$; the point of contact between the upper and lower portions of
the string  slides horizontally without friction along the line
$CD$. The pulleys at $A$ and $B$ are considered frictionless and
with zero inertia; therefore the tensions of the different
portions of the string will be $T_1$ and $T_2$. Compare the
potential energy of the configurations where the point of contact
are $E$ and $e$. There is a  potential energy difference between
the two configurations because in going from $E$ to $e$, mass 1
will rise from $M$ to $m$ and  mass 2 will decrease its height
from $N$ to $n$. The tensions are determined by the weights and
are, therefore, in both configurations, $T_1$ and $T_2$. Calling
$\ell _1= AE$ and $\ell _2= EB$, the change in potential energy
$\Delta U$ of the system becomes
\begin{equation}
\Delta U= T_1 \Delta \ell _1+ T_2 \Delta \ell _2, \label{ustring}
\end{equation}
where $\Delta \ell _1=Mm\equiv qe$ and $\Delta \ell _2=-Nn\equiv
Ep$. This means that, up to an additive constant, the potential
energy of the system can be written as
\begin{equation}
 U= T_1 \ell _1+ T_2  \ell _2. \label{ustring2}
\end{equation}

\begin{figure}
\includegraphics*[width=0.3\textwidth]{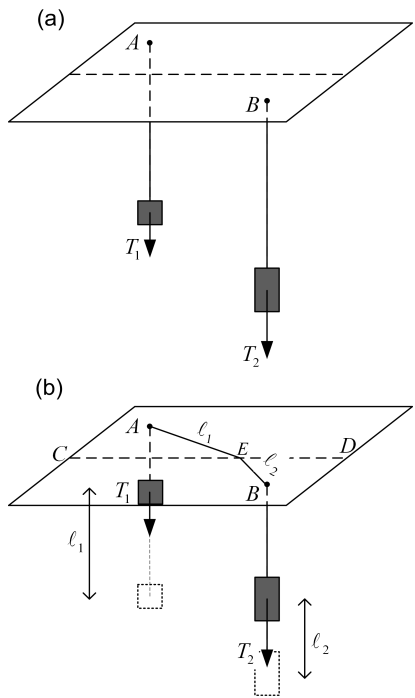}
\vspace{-0.3cm}

 \caption{An alternative version of Bernoulli's setup: (a)
 Two weights hanging
 from frictionless points $A$ and $B$ is assigned zero potential energy.
  (b) The two weights are then lifted, and the ends
 of the strings are joined at point $E$ along the line $CD$. The work done is
 equal to the increase in potential energy: $U=T_1\ell_1+T_2\ell_2$.}
 \label{table}
\end{figure}

Another way of visualizing the above expression is offered in
Figure \ref{table}.

Since the configuration of  mechanical equilibrium corresponds to
the minimum of potential energy  the minimum $U$ is attained when
the components of the forces from the different portions of the
string along $CD$ cancel.  In term of the angles $\angle AER=
\theta_1 $ ,
 $\angle SEB= \theta_2 $, the minimum potential energy is attained
 when
\begin{equation}
T_1\sin \theta_1=T_2\sin \theta_2.
\end{equation}

The above relation is equivalent to Snel's law if  the indices of
refraction $n_1$ and $n_2$ in the different regions are identified
with the corresponding tension of the strings: if  the time $t$ is
minimized, or equivalently the optical length $ct$ given by
\begin{equation}
ct=n_1\ell_1 +n_2\ell _2, \label{ct}
\end{equation}
then  $n_1\sin \theta_1=n_2\sin \theta_2$. For the analogy between
the configuration of the tense string and the trajectory of a
particle with energy $E$ between points $A$ and $B$
 I reason in reverse: knowing that conservation of momentum
 at the interface requires
 \begin{equation}
 mv_1\sin\theta_1=mv_2\sin \theta_2,
 \label{snellp}
 \end{equation}
what is the magnitude that should be minimized in order to get the
above relation? Clearly, from the above analysis the quantity to
be minimized is
\begin{equation}
A=mv_1\ell_1+mv_2\ell_2,
\label{maup}
\end{equation}
which is Maupertuis' action. The correspondence between Eq.
\ref{ct} and Eq. \ref{maup} expresses the famous analogy between
mechanics and geometric optics that has been  the subject of many
recent pedagogical expositions\cite{fma,pform}.

\begin{figure}
\includegraphics*[width=0.25\textwidth]{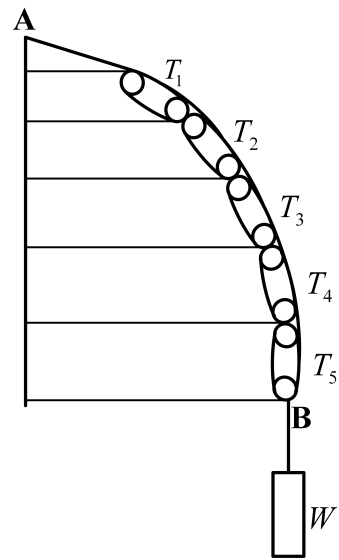}
\vspace{-.0cm} \caption{Frictionless pulleys that can slide in
horizontal lines with a string passing through them a sufficient
number of times gives the trajectory of the particle if $T_i$ is
identified with $mv_i$ at each segment. Since the string can only
pass through each pulley an integer number of times, the ratios of
the velocities are approximated by the ratio of the times the rope
passes through each segment.}  \label{figmach}
\end{figure}

The tension $T_i$ of the string is identified with the velocity
$v_i$ on each region which, in turn, is given by
$v_i=\sqrt{2m(E-U_i)}$, where $E$ is the total energy and  $U_i$
is the corresponding potential energy. For paths that transverse
many regions where the particle velocities are different,  the
trajectory has to be broken up in many straight segments. The
Principle of Least Action states that the trajectory the particle
will follow between two fixed points and at fixed total energy is
the one that minimizes the sum of the products $mv_i\ell_i$ in
each segment. In order to use the analogy with the string, the
arrangement will correspond to frictionless pulleys that can slide
on rods, with the string passing  through them a sufficient number
of
 times\cite{mach} (see Figure \ref{figmach}).

Table \ref{analogy}  summarizes the analogies between the
quantities discussed in this section.

\begin{table*}
\caption{\label{analogy} Corresponding quantities in the analogy
used in the Principle of Least Action between mechanics, geometric
optics and the equilibrium of a non-stretchable string. }
\begin{ruledtabular}
\begin{tabular}{ccc}
Particle &Light Ray&Non-stretchable string \\
\hline
$mv$ (momentum)& $n$ (refractive index)  & $T$ (tension)  \\
$mv_1\sin \theta_1=mv_2\sin\theta_2$ (momentum conservation)&  $n_1\sin \theta_1=n_2\sin\theta_2$
(Snel's law)& $T_1\sin \theta_1=T_2\sin\theta_2$ (static equilibrium) \\
  $\Delta A=mv\Delta \ell$ (action) & $c\Delta t=n\Delta \ell$ (optical length) & $\Delta U=T\Delta \ell$ (potential energy)  \\
\end{tabular}
\end{ruledtabular}
\end{table*}

\section{\label{hamnc} Hamilton's principle and stretchable strings}

The Principle of Least Action as stated in the previous paragraph gives
 the optimal trajectory for a particle of a given energy between two fixed points.
It doesn't say anything about the time it takes to travel from one
point to the other. Now consider the problem of finding a path
that will connect point $P$ to point $Q$ in a fixed time $t$.
 In other words, what is  the path that
the particle will choose in going from point ${\rm{\bf
P}}=(x_P,0)$ to point ${\rm{\bf{Q}}}=(x_Q,t)$? In extending the
treatment to paths that go from two fixed space--time points it is
useful to treat $t$ as a new geometrical dimension. To simplify
the analysis, and to retain the two dimensional picture of the
previous paragraph, I will consider motion in one (spatial)
 dimension.

\begin{figure}
\includegraphics*[width=0.3\textwidth]{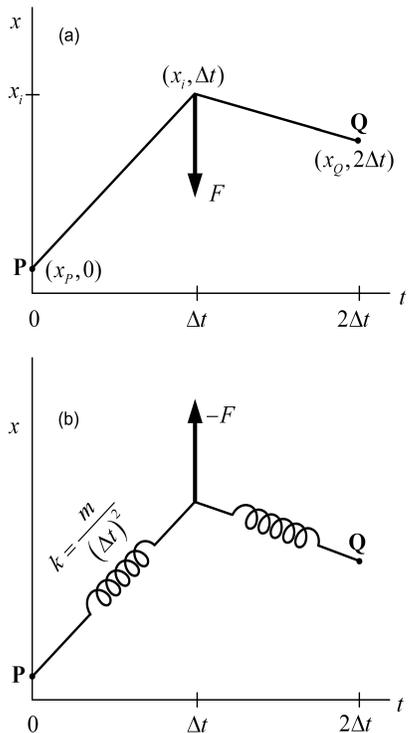}
\caption{(a) Space-time trajectory of an otherwise free one-dimensional particle acted
upon by an impulsive force $F$ at $t_i=\Delta t$. (b) Equivalent equilibrium configuration
of two segments of a  stretchable string with spring constants $k=m/(\Delta t)^2$ and an
external force $-F$.}
\label{xvst}
\end{figure}

 Consider the path as a  continuous function
 $x(t)$,
 that is broken into small straight segments connecting points separated
  by a fixed time interval $\Delta t$. The fact that the segments are straight means
   that the motion is of constant velocity during that interval,   then
   changes  due to an impulsive force. This force will be different
   from zero if the potential is changing as a function of $x$ at the particle
   position. Consider for simplicity two broken segments as in Figure \ref{xvst}(a).

Before the force $F$ acts on the particle, the velocity
  is given by the slope of the curve in the $(x,t)$ plot:
\begin{equation}
         v_P={x_i-x_P\over\Delta t}.
\end{equation}

The effect of the force is to change  the velocity. In the $(x,t)$
plot (also called the ``world line"), this means that the slope of
the
 line changes at the intermediate time. The velocity after the force has acted
 on the particle is

  \begin{equation}
           v_Q={x_Q-x_i\over\Delta t}.
  \end{equation}

Notice that, since the force is downward, the slope decreases:
downward force means
   that, at $x_i$ , the potential is increasing as a function of $x$.

The path
 [in space time $(x,t)$] is a solution
 of Newton's second law, according to which
  the rate of change of the velocity times the
  particle mass is the force acting on it:

   \begin{equation}
            F=m{v_Q-v_P\over\Delta t}.
   \end{equation}

Replacing the above expressions for the velocity:

\begin{equation}
                   F={m\over \left(\Delta t\right)^2} \left(x_Q-x_i\right)
                   -  {m\over \left(\Delta t\right)^2}
                   \left(x_i-x_P\right).
\end{equation}

At this point   take a step of abstraction, and forget the
space-time picture for a moment. This expression gives the force
of a system of two springs of identical spring constant,
 the first spring connecting point  $(x_i,\Delta t)$ with $(x_P,0)$,
the second connecting $(x_Q,2\Delta t)$  with $(x_i,\Delta t)$. In
order for the systems to be in equilibrium, or, in other words,
for the  intermediate coordinate to have the value $x_i$ (the
other two are fixed) there has to be a force of precisely
magnitude $F$ but of {\em opposite} sign.

The  path given by Newton's law  is given by the equilibrium condition
 of a mechanical model of two springs in the presence
  of a potential of opposite sign as $U(x)$. The equilibrium
   condition is the one that minimizes the  potential
   energy of the entire system, springs plus ``external"
   potential $U(x)$. Since the potential energy for a spring
   of spring constant $k$ connecting two points separated by
    a distance $\delta$  is $k\delta ^2/2$ , the total potential
    energy of the system (that I will call $\widetilde{S}$) is
    given by
\begin{equation}
   \widetilde{S}=
          {m\over 2}\left( {x_i-x_P \over \Delta t}\right)^2
          +
{m\over 2}\left( {x_Q-x_i \over \Delta t}\right)^2 -U(x_i).
\label{s1}
\end{equation}

Coming back to the original world line picture, the optimum path
in space--time is the one that
 minimizes the difference between kinetic and
 potential energy. This is Hamilton's principle,
  which I obtained using a mechanical analogy
  similar to the Principle of Least Action in
  the sense that there is a correspondence between the
   kinetic energy and the potential energy of
    fictitious springs of spring constant
$k=m/(\Delta t)^2$. In other words, the stretchable string is in
equilibrium due to two types of forces in space-time: the external
force due to  (minus) the real external potential, and the elastic
force of fictitious springs playing the role of the kinetic
energy.

For a   a longer path with $N$ straight segments each of  them
traversed by the particle in a time $\Delta t$, the velocity at
the $i$-th segment will be $v_i=(x_{i+1}-x_i)/\Delta t$ and the
equivalent potential energy  will be given by

\begin{equation}
   \widetilde{S}=
          \left[{mv_1^2\over 2}-U(x_1)\right]
+ \left[{mv_2^2\over 2}-U(x_2)\right] + \cdots +{mv_N^2\over 2}.
\label{s2}
\end{equation}

For a large number of segments, which will correspond to a
continuously varying path, the last term in the sum above can be
ignored; the total potential energy to be minimized is the sum
 of the differences between the kinetic and potential energies.

 Notice that the potential energy for the fictitious springs
 corresponds to springs of {\em zero length}. Also,  Equations \ref{s1} and \ref{s2}
 omit the potential energy associated with the ``horizontal displacement" $\Delta t$ of
 each spring. I ignore this contribution because the horizontal forces due to the springs
  cancel and, therefore, the potential energy associated with this displacement is the same for
  all configurations of the string.

    \section{\label{hamc1} Hamilton's principle using elementary calculus}

\subsection{Connection between Maupertuis' and Hamilton's action}

 In this section I, derive Hamilton's principle using a
slightly more sophisticated approach while still keeping it
elementary. In reference to  Figure \ref{figber}(a), the Principle
of Least Action tells us  the path  a particle of fixed energy $E$
will choose in going from $A$ to $B$. Call $U_1$ and $U_2$ the
potential energies in the upper and lower parts of the line $CD$,
and $v_1=\sqrt{2m(E-U_1)}$ and $v_2=\sqrt{2m(E-U_2)}$ the
corresponding velocities. Now consider paths with different
energies and ask for which of those paths the particle will
satisfy Newton's laws and spend a fixed amount of time $t$  going
from $A$ to $B$.  Following the logic of the Principle of Least
action, I want to find a function of the paths that will give the
desired one upon minimization. For the special case in
consideration, the path consists of two straight segments and the
function  has to be such that, of all paths that take a time $t$
in going from $A$ to $B$, it chooses the one that satisfies  the
``Snel's law for particles" of Eq. \ref{snellp}.

Call $a$ and $b$ the perpendicular distances of $A$ and $B$ to the
interface $CD$, $L$ the horizontal distance between $A$ and $B$
and $x$ the distance $CE$.
 Maupertius action of Eq. \ref{maup}
can be thought of as a function of $x$ and the energy:
\begin{equation}
A(x,E)=mv_1(E) \sqrt{x^2+a^2}+mv_2(E)\sqrt{(L-x)^2+a^2}.
\end{equation}

In order to explore whether $A(x,E)$ is the desired function,
compute the variations of $A$ with respect to $x$ and $E$,
assuming knowledge of the ratios of $dE$ and $dx$ that will keep
the time $t$ constant:
\begin{equation}
dA=\left(m v_1\sin \theta_1 - m v_2 \sin \theta_2\right)dx
+{\partial A\over \partial E} dE.
\end{equation}

It is clear that minimizing $A$ (or equivalently setting $dA=0$)
does not give us the desired Eq.\ref{snellp} because of the second
term above. However,  notice that $d v_i/dE=1/mv_i$, and
\begin{equation}
{\partial A\over \partial E}={\sqrt{x^2+a^2}\over v_1}+{\sqrt{(L-x)^2+a^2}\over v_2}=
t_1+t_2=t,
\end{equation}
with $t_1\equiv \ell_1/v_1$ and $t_2\equiv \ell_2/v_2$  the times it takes the particle to go
from $A$ to $B$ and  from $E$ to $B$.

 This means that if   $Et$ is {\em subtracted} from $A$  the desired quantity is obtained: $S=A-Et$. (Notice
that $d(Et)=tdE$ since the paths considered   last a constant
time.) Therefore,
 \begin{eqnarray}
 S&=&\left(mv_1\ell_1-Et_1\right)+
  \left(mv_2\ell_1-Et_2\right)
\nonumber \\
  \nonumber
  &=&\left(mv_1^2-E\right)t_1+
  \left(mv_2^2-E\right)t_2\\
   &=&\left(K_1-U_1\right)t_1+
  \left(K_2-U_2\right)t_2,
  \label{ss}
 \end{eqnarray}
 which is the quantity to be minimized according to Hamilton's principle.

  \subsection{\label{hamq} Could Hamilton have discovered quantum mechanics in 1834?}
Writing  $p=mv$ and $\ell=x$, the above action
 $S$ (Eq.\ref{ss}) for a path has precisely the form of the phase change of a
 wave
 \begin{equation}\phi \sim px-Et
\label{phase}
 \end{equation}
 (up to a multiplicative constant that will
 render it dimension-less) with momentum and energy playing the role of wave number $k$ and frequency $\omega$.
  The path of least--or, more precisely
 stationary--action could hence be regarded as the stationary phase
 limit of
 some wave.
 With this motivation I ask wether Hamilton could have discovered quantum mechanics in 1834.
 The answer  is most probably ``no," since Hamilton did not have any
 experimental motivation to think of particles as waves.\cite{hamilton0} However, the close parallelism
 between geometric optics and mechanics could have invited him to ask himself the following: what
 would be the structure of a wave equation for particles that, in the limit of small
 wavelengths,
 gives the  trajectories of particles just as the wave equation for light in the same limit gives
 the trajectories of light rays?

Let us follow the analogy provided by the Principle of Least
Action and consider trajectories of constant energy for particles
which will correspond to light rays of constant frequency. Since
the Principle of Least Action establishes an equivalence between
the geometry of trajectories and not between the dynamics of those
trajectories, for constant energy (and frequency) I seek an
equivalence between stationary states of the corresponding wave
equations.

Now let us introduce wave-lengths into the discussion. The wave length of
a monochromatic light wave in a region in which the
 index of refraction $n(x)$ is varying slowly is given by

 \begin{equation}
 \lambda(x) ={\lambda _0\over n(x)}.
\label{q1}
\end{equation}

Since equations \ref{ct} and \ref{maup} imply that  the
trajectories of particles and light rays are equivalent if $n(x)$
is identified with $mv(x)$, the ``natural" choice for the spatial
dependence of the particle wave length $\lambda _P$ is

 \begin{equation}
 \lambda_P(x) ={K\over mv(x)}={K\over \sqrt{2m\left[E-U(x)\right]}},
\label{q2}
\end{equation}
 with $K$ a constant that has units of angular momentum and
 is precisely the required constant to make $\phi$ in Eq. \ref{phase}
 dimension-less:
 $\phi={2 \pi\over K}(px-Et)$.

 Now let us consider the wave equation for the  amplitude $\phi(x,t)$ describing
 a light wave in one dimension (I ignore the polarization
 and consider it as a scalar wave)\cite{foot1}:
 \begin{equation}
 {\partial^2 \phi\over \partial x^2}={n^2(x)\over c^2}{\partial^2 \phi  \over \partial t^2}.
 \label{wave1}
 \end{equation}

 I want to compare this with the stationary wave equation for particles, so I substitute
$\phi(x,t)=\phi(x)e^{i\omega t}$

and the equation becomes
 \begin{equation}
 -\left({\lambda_0 \over 2\pi }\right)^2{\partial^2 \phi\over \partial x^2}=n^2(x)
 \phi  .
 \label{light}
 \end{equation}

Using the equivalences of Eqs. \ref{q1} and \ref{q2}, the
structure of the wave equation for the stationary states $\Psi$
for particles should be
\begin{equation}
 -{\left(K/2\pi\right)^2\over 2m}{\partial^2 \Psi\over \partial x^2}=\left[ E-U(x) \right]
 \Psi.
 \label{sch}
 \end{equation}

 In  comparison with Eq. \ref{light},  if  $K$ is treated as
a free parameter, the limit $K\rightarrow 0$ (which corresponds to
the limit $\lambda_0 \rightarrow 0$)
 gives  the trajectories for particles of energy $E$
   and  Eq. \ref{sch}
could be used as a wave equation for particles. If many years
later someone were to discover that particles in confined
potentials had discrete energies,   $K$ could be tuned to fit the
experiments obtaining
\begin{equation}
{K\over 2\pi}=1.0546\times 10^{-34}{\rm J\, s},
\end{equation}
which is of course Planck's constant $\hbar$. Given the
identification of energy with frequency, the  time dependence of
the stationary states should be $\Psi(x) e^{-iEt/\hbar}$ and the
``naturally" implied time dependence for Eq. \ref{sch} is
\begin{equation}
 \left[-{\hbar^2\over 2m}{\partial^2 \over \partial
 x^2}+U(x)\right]\Psi=i\hbar {\partial \over \partial t}
\Psi,
\end{equation}
which is the celebrated Schr\"{o}dinger equation.

\section{conclusion}

I have presented an elementary derivation of Hamilton's principle
based on the analogy between particle (and light ray) trajectories
with the configuration of a tense non-stretchable string. The
analogy was (to my knowledge) first considered by Bernoulli in a
practically unreferenced article and gives an intuitive picture of
the Principle of Least Action. Extending the analogy to
stretchable strings, I derived Hamilton's principle,  providing an
intuitive picture of the origin of the difference between kinetic
and potential energy in Hamilton's characteristic function. I also
presented a derivation using elementary calculus, which  extends
the analogy between mechanics and geometrical optics to undulatory
mechanics and ``derived" Schr\"{o}dinger's equation.

I have endeavored to present a pedagogical approach that can help
students in introductory courses appreciate the beauty and
compactness of the   Principle of Least Action.

\begin{acknowledgments}
I wish to thank Anthony Bloch, Roberto Rojo  and Arturo L\'{o}pez
D\'{a}valos for conversations,  Alejandro Garc\'{\i}a, Paul Berman
and an anonymous referee for valuable corrections, and Estela
As\'{\i}s for her help with the translation from the Latin of
Ref.\cite{bernoulli}. This work is partially supported by Research
Corporation, Cotrell College Science Award.
\end{acknowledgments}


\begin{thebibliography}{9}
\bibitem{foot0} Strictly speaking,   extremal (maxima or minima) are a subset
 stationary paths (or states). However, the term Principle of Least Action is
 more frequently used than the more precise Principle of
 Stationary
 Action.
\bibitem{foot1} In  consonance with this historical excercise it could be objected that Hamilton
 didn't know the wave equation for light, but the argument applies to any scalar wave in a nonhomogeneous
 medium. For example, Eq. \ref{wave1} could represent a sound wave in a medium with varying density.
\bibitem{hamilton0}The title of this section was taken from
 Ref.\cite{hamilton1} p. 209.
\end{thebibliography}
\end{document}